\documentclass[a4paper,aps,onecolumn,nofootinbib]{revtex4}
\RequirePackage[colorlinks,hyperindex]{hyperref}
\RequirePackage[english]{babel}
\RequirePackage[latin1]{inputenc}
\RequirePackage[T1]{fontenc}
\RequirePackage{mathrsfs}
\RequirePackage{amsmath}
\RequirePackage{amssymb}
\RequirePackage{amsbsy}
\RequirePackage{color}
\RequirePackage{bm}
\hypersetup{colorlinks=true,breaklinks=true,urlcolor=blue,linkcolor=red}
\newfont{\gotico}{eufm10 scaled\magstephalf}
\newfont{\qvd}{msam10 scaled\magstephalf}

\def\de#1/de#2{\frac{\partial {#1}}{\partial {#2}}}
\def\De#1/de#2{\dfrac{\partial {#1}}{\partial {#2}}}
\begin{document}
\title{Dirac spinors and their application to Bianchi-I space-times in 5 dimensions}
\author{Stefano Vignolo}
\email{vignolo@dime.unige.it}
\author{Luca Fabbri}
\email{fabbri@dime.unige.it}
\affiliation{DIME Sez. Metodi e Modelli Matematici, Universit\`{a} di Genova, Via all'Opera Pia 15, 16145 Genova ITALY.}
\author{Oscar Castillo-Felisola}
\email{o.castillo.felisola@gmail.com}
\affiliation{Departamento de F\'\i sica, Universidad T\'ecnica Federico Santa Mar\'\i a, Casilla 110-V, Valpara\'\i so, and\\ Centro Cient\'ifico Tecnol\'ogico de Valpara\'iso, Chile.}
\date{\today}
\begin{abstract}
We consider a five-dimensional Einstein-Sciama-Kibble spacetime upon which Dirac spinor fields can be defined. Dirac spinor fields in five and four dimensions share many features, like the fact that both are described by four-component spinor fields, but they are also characterized by strong differences, like the fact that in five dimensions we do not have the possibility to project on left-handed and right-handed chiral parts: we conduct a polar decomposition of the spinorial fields, so to highlight all similarities and discrepancies. As an application of spinor fields in five dimensions, we study Bianchi-I spacetimes, verifying whether the Dirac fields in five dimensions can give rise to inflation or dark-energy dominated cosmological eras or not.
\end{abstract}
\maketitle
\section{Introduction}
The spin-$\frac{1}{2}$ spinor field is the only type of spinor field that we have observed so far in nature and quite possibly one of the most fundamental fields we can define in general: according to the well-known Lounesto classification \cite{L}, it may represent a regular spinor, like a Dirac field (describing massive-charged particles), or it may represent a singular spinor, either of Weyl type (describing massless particles) or of Majorana type (describing neutral particles).

Particles such as the neutrinos may be Majorana fields, while charged leptons and quarks are Weyl or Dirac according to whether they are considered before or after symmetry breaking in the standard model of particle physics.

One key property of all these types of spinor fields is the fact that they are rather sensitive to the dimension of the space in which they live: so for instance, in three dimensions, Dirac spinors are the well known Pauli spinors, and they are described by a two components complex (column) fields; in four dimensions, however, Dirac spinors are described by a four components complex (column) fields. The differences arise from the fact that Dirac spinor fields are defined in terms of an underlying structure known as Clifford algebra, and it is this algebra that is sensitive to the dimension: so for example, in three dimensions the Clifford algebra is built up in terms of three mutually anti-commuting matrices that can be taken to be the $2\times2$ Pauli matrices, while in four dimensions the Clifford algebra must contain a fourth matrix anti-commuting with the other three matrices and there is no way to do this unless the Pauli matrices are extended to the $4\times4$ Dirac matrices. Hence, in three and four dimensions, the corresponding spinor fields have two and four complex components, respectively.

Nevertheless, spinor fields defined in a given odd dimension preserves the spinorial structure of the co-dimension one spacetime, i.e., spinor fields defined in two dimensions have the same number of components as the spinor fields defined in three dimensions, as well as spinor fields defined four dimensions have the same number of components as the spinor fields defined in five dimensions. And therefore, taking the standard definition of spinor field as the four-dimensional one, the
five-dimensional space is somewhat special, because among all different spacetimes, it is the only one for which the spinorial structure is unchanged.

As a consequence, it becomes interesting to ask what would change in the spinorial structure if the four-dimensional spinor were not defined in four dimensions but obtained as the result of a dimensional reduction from a higher-dimensional spacetime, and among all of them the five-dimensional spacetime is a perfect starting place.

In this paper we are going to do precisely this: we will give the five-dimensional definition of spinors and eventually reduce to the four-dimensional standard case, and in the process we shall stress on the analogies and differences between the two approaches.

As an application of our study, we investigate cosmological scenarios arising from the presence of Dirac fields in five-dimensional Bianchi-I Universes. In cosmology, spinor fields have been largely
studied, both minimally and non-minimally coupled to gravity (for example, see references \cite{Saha1,Saha3,Saha2,Saha,Saha4, Ribas1,Souza,Ribas, Grams,CVC,VCF} and references therein); in general, non-minimal coupling or self-interaction potentials are seen to be necessary for
the Dirac fields to generate inflationary or dark-energy dominated eras. Here we consider Dirac fields minimally coupled to five-dimensional Einstein-Cartan gravity, without self-interaction
potential. More in particular, relying on the additional dimensional degree of freedom, we want to verify whether a contraction in the fourth spatial dimension corresponds to a (possibly accelerated) expansion in the remaining other three dimensions, thus giving rise to
a four-dimensional expanding Universe. As we shall see, the non-diagonal part of the Einstein-like equations, together with the anisotropy on the fourth spatial dimension, impose stricter constraints than those appearing in four dimensions. This reduces the admissible
forms of the spinor fields and simplifies the resulting field equations. We give explicit solutions of the dynamical equations, showing that Dirac fields in five dimensions can not be considered as
the source of accelerated expansions of our Universe, because at the most they are seen to generate Friedmann eras.
\section{Geometrical preliminaries of the five-dimensional space}
\subsection{Kinematic structures}
Let us consider a five-dimensional manifold $\pi: Q \to M$, fibered over a four-dimensional spacetime $M$, and allowing a metric tensor of signature $(1,-1,-1,-1,-1)$. For our purposes, we do not need to assume that the fibration $\pi: Q \to M$ has any particular structure (for example, we do not require $\pi: Q \to M$ to be a principal fiber bundle).

Introducing a f\"{u}nfbein $e^\mu=e^\mu_i\,dx^i$ defined on the manifold $Q$, the metric tensor can be expressed as $g=\eta_{\mu\nu}\,e^\mu\otimes e^\nu$ where $\eta_{\mu\nu}=\eta^{\mu\nu}={\rm diag}(1,-1,-1,-1,-1)$. Greek and Latin indices run from zero to four: Latin indices label local coordinates on the manifold $Q$, while Greek indices label elements of local orthonormal frames and co-frames undergoing five-dimensional Lorentz transformations. Writing $e_\mu=e^i_\mu\partial_{i}$ as the dual frame of $e^\mu$ we have
\begin{subequations}
\label{duality}
\begin{eqnarray}
e^j_\mu e^\mu_i= \delta^j_i\\
e^j_\mu e^\nu_j= \delta^\nu_\mu 
\end{eqnarray}
\end{subequations}
as the duality relations allowing us to get the f\"{u}nfbein from the dual f\"{u}nfbein.

The assignment of a metric-compatible linear connection $\Gamma_{ij}{}^{h}$ on $Q$ induces a corresponding spin-connection defined as
\begin{equation}
 \label{spinconnection}
 \omega^{\;\;\mu}_{i\;\;\;\nu} = \Gamma_{ij}^{\;\;\;h} e^\mu_h e^j_\nu - e^j_\nu \de{e^\mu_j}/de{x^i}\ \ \ \ \mathrm{with}\ \ \ \ \omega_i^{\;\;\mu\nu} =-\omega_i^{\;\;\nu\mu}
\end{equation} 
according to the requirement that the covariant derivative applied to the f\"{u}nfbein and the metric be zero identically.

The simultaneous introduction of a f\"{u}nfbein and a spin-connection generates corresponding torsion and curvature tensors expressed in local coordinates as
\begin{subequations}
 \label{tensors}
 \begin{equation}
 \label{torsion}
 T^{\;\;\mu}_{ij} =\partial_i e^\mu_j-\partial_j e^\mu_i+\omega^{\;\;\mu}_{i\;\;\;\nu}e^\nu_{j} 
 -\omega^{\;\;\mu}_{j\;\;\;\nu}e^\nu_{i}
 \end{equation}
 \begin{equation}
 \label{curvature}
 R_{ij}^{\;\;\;\;\mu\nu} =\partial_i\omega_{j}^{\;\;\mu\nu}-\partial_j{\omega_{i}^{\;\;\mu\nu}}
 +\omega^{\;\;\mu}_{i\;\;\;\lambda}\omega_{j}^{\;\;\lambda\nu}
 -\omega^{\;\;\mu}_{j\;\;\;\lambda}\omega_{i}^{\;\;\lambda\nu}
 \end{equation}
\end{subequations}
and by contraction, from equation \eqref{curvature} we derive the expressions of the Ricci tensor $R^i_{\;j}:=R_{\sigma\mu}^{\phantom{\sigma\mu}\sigma\lambda}e^{i}_{\lambda}e^\mu_j$ and the Ricci scalar $R= R_{ij}^{\phantom{ij}\mu\nu} e^i_\mu e^j_\nu$, while no contraction will be considered for torsion since it will be taken to be completely antisymmetric (the reasons of this will become clear later on in the development of the theory).

Following a standard procedure, we decompose the linear connection $\Gamma_{ij}^{\;\;\;h}$ into the Levi-Civita connection $\tilde{\Gamma}_{ij}^{\;\;\;h}$ (associated with the metric $g$) plus torsional contributions so that
\begin{equation}
 \label{covariantderivative}
 \nabla_{i}A_{j} = \tilde{\nabla}_{i}A_{j} - \frac{1}{2}T^{\phantom{ij}h}_{ij}A_{h}
\end{equation}
where $\tilde{\nabla}_{i}$ is the Levi-Civita covariant derivative and where the total antisymmetry of torsion was used. Analogously, given the antisymmetry of torsion, the Ricci tensor and Ricci scalar are decomposed as
\begin{subequations}
 \label{tensordecomposition}
 \begin{equation}
 R_{ij}=\tilde{R}_{ij}+\frac{1}{2} \tilde{\nabla}_p T_{ji}^{\;\;\;p}
 -\frac{1}{4}T_{pi}^{\;\;\;q} T_{jq}^{\;\;\;p}
 \end{equation}
 \begin{equation}
 R=\tilde{R}-\frac{1}{4}T_{qpr}T^{qpr}
 \end{equation}
\end{subequations}
with $\tilde{R}_{ij}$ and $\tilde{R}$ respectively the Ricci tensor and Ricci scalar of the spin connection $\tilde{\omega}_{j}^{\;\;\mu\nu}$ which is associated with the Levi-Civita connection.

For the matter content, we employ representations of the Clifford algebra given in terms of five Dirac matrices $\gamma^{\mu}$ satisfying
\begin{equation}
 \left\{\gamma_{\mu},\gamma_{\nu}\right\} = 2 \eta_{\mu\nu} \mathbb{I}
\end{equation}
like in the four-dimensional case. We recall that in four as well as five dimensions, the minimal size of these Dirac matrices is $4\times 4$ although in the five-dimensional situation they will no longer be block-diagonal, even in chiral representation, as we shall see in a while. Defining
\begin{equation}
 S_{\mu\nu} = \frac{1}{8} \left[\gamma_{\mu},\gamma_{\nu}\right]
\end{equation}
it is possible to verify that they satisfy the commutation relationships of the Lorentz algebra, although in five dimensions there no longer is a non-identity matrix commuting with all generators: this shows that no representation can be reducible, compatibly with the fact that we cannot find block-diagonal representations. We also have the identities
\begin{equation}
 \gamma_{\mu}\gamma_{\nu}\gamma_{\lambda} = \eta_{\nu\lambda}\gamma_{\mu}
 -\eta_{\mu\lambda}\gamma_{\nu}+\eta_{\mu\nu}\gamma_{\lambda}
 -\frac{1}{2}\epsilon_{\mu\nu\lambda\alpha\beta}\gamma^{\alpha}\gamma^{\beta}
\end{equation}
as well as the contractions
\begin{eqnarray}
&\gamma^{\mu}S^{\alpha\beta}\gamma_{\mu}= S^{\alpha\beta}\\
&\{\gamma^\mu,S^{\nu\lambda}\}= -\epsilon^{\mu\nu\lambda\tau\rho}S_{\tau\rho}\\
&[\gamma^\mu,S^{\nu\lambda}]= \frac{1}{2}(\gamma^\lambda\eta^{\mu\nu}-\gamma^\nu\eta^{\mu\lambda})\label{comm}
\end{eqnarray}
with $\epsilon^{\mu\nu\lambda\tau\rho}$ denoting the Levi-Civita completely antisymmetric five-dimensional pseudo-tensor. It is important to notice that the introduction of this tensor makes the set of formulas sensitive to the dimension of the space, as for example the first two of these identities would be very different for the standard number of four dimensions, although equation \eqref{comm} remains unchanged, as it should be, since it is what ensures the correct transformation law of the Dirac matrices, and so the Lorentzian structure.

An explicit choice of the Dirac matrices is the chiral representation
\begin{equation}
 \begin{pmatrix}
 0 & \mathbb{I}
 \\
 \mathbb{I} & 0
 \end{pmatrix} = \gamma^{0}
 \quad
 \begin{pmatrix}
 0 & \sigma^\mathcal{A}
 \\
 -\sigma^\mathcal{A} & 0
 \end{pmatrix} = \gamma^\mathcal{A}
 \quad 
 \begin{pmatrix}
 i\mathbb{I} & 0\\
 0 &-i\mathbb{I}
 \end{pmatrix} = \gamma^{4}
\end{equation}
where the sigmas ($\mathcal{A}=1,2,3$) are the Pauli matrices.

The complex Lorentz transformation laws are given by $\boldsymbol{\Lambda} = e^{S^{ab}\theta_{ab}}$ and they amount to
\begin{equation}
 \begin{aligned}
 \boldsymbol{\Lambda}_{R12}= \begin{pmatrix}
 e^{i\frac{\theta}{2}} & 0 & 0 & 0 \\ 
 0 & e^{-i\frac{\theta}{2}} & 0 & 0 \\ 
 0 & 0 & e^{i\frac{\theta}{2}} & 0 \\ 
 0 & 0 & 0 & e^{-i\frac{\theta}{2}}
 \end{pmatrix}
 &
 \boldsymbol{\Lambda}_{R34}= \begin{pmatrix}
 \cos{\frac{\theta}{2}} & 0 & i\sin{\frac{\theta}{2}} & 0 \\ 
 0 & \cos{\frac{\theta}{2}} & 0 & -i\sin{\frac{\theta}{2}} \\ 
 i\sin{\frac{\theta}{2}} & 0 & \cos{\frac{\theta}{2}} & 0 \\ 
 0 & -i\sin{\frac{\theta}{2}} & 0 & \cos{\frac{\theta}{2}}
 \end{pmatrix}
 \\
 \boldsymbol{\Lambda}_{R31}= \begin{pmatrix}
 \cos{\frac{\theta}{2}} & \sin{\frac{\theta}{2}} & 0 & 0 \\ 
 -\sin{\frac{\theta}{2}} & \cos{\frac{\theta}{2}} & 0 & 0 \\ 
 0 & 0 & \cos{\frac{\theta}{2}} & \sin{\frac{\theta}{2}} \\ 
 0 & 0 & -\sin{\frac{\theta}{2}} & \cos{\frac{\theta}{2}}
 \end{pmatrix}
 &
 \boldsymbol{\Lambda}_{R42}= \begin{pmatrix}
 \cos{\frac{\theta}{2}} & 0 & 0 & \sin{\frac{\theta}{2}} \\ 
 0 & \cos{\frac{\theta}{2}} & -\sin{\frac{\theta}{2}} & 0 \\ 
 0 & \sin{\frac{\theta}{2}} & \cos{\frac{\theta}{2}} & 0 \\ 
 -\sin{\frac{\theta}{2}} & 0 & 0 & \cos{\frac{\theta}{2}}
 \end{pmatrix}
 \\
 \boldsymbol{\Lambda}_{R23}= \begin{pmatrix}
 \cos{\frac{\theta}{2}} & i\sin{\frac{\theta}{2}} & 0 & 0 \\ 
 i\sin{\frac{\theta}{2}} & \cos{\frac{\theta}{2}} & 0 & 0 \\ 
 0 & 0 & \cos{\frac{\theta}{2}} & i\sin{\frac{\theta}{2}} \\ 
 0 & 0 & i\sin{\frac{\theta}{2}} & \cos{\frac{\theta}{2}}
 \end{pmatrix}
 &
 \boldsymbol{\Lambda}_{R14}= \begin{pmatrix}
 \cos{\frac{\theta}{2}} & 0 & 0 & i\sin{\frac{\theta}{2}} \\ 
 0 & \cos{\frac{\theta}{2}} & i\sin{\frac{\theta}{2}} & 0 \\ 
 0 & i\sin{\frac{\theta}{2}} & \cos{\frac{\theta}{2}} & 0 \\ 
 i\sin{\frac{\theta}{2}} & 0 & 0 & \cos{\frac{\theta}{2}}
 \end{pmatrix}
 \end{aligned}
\end{equation}
as the rotations in all possible pairs of planes with $\theta$ angles and 
\begin{equation}
 \begin{aligned}
 \boldsymbol{\Lambda}_{B1}= \begin{pmatrix}
 \cosh{\frac{\zeta}{2}} & \sinh{\frac{\zeta}{2}} & 0 & 0 \\ 
 \sinh{\frac{\zeta}{2}} & \cosh{\frac{\zeta}{2}} & 0 & 0 \\ 
 0 & 0 & \cosh{\frac{\zeta}{2}} & -\sinh{\frac{\zeta}{2}} \\ 
 0 & 0 & -\sinh{\frac{\zeta}{2}} & \cosh{\frac{\zeta}{2}}
 \end{pmatrix}
 &
 \boldsymbol{\Lambda}_{B2}= \begin{pmatrix}
 \cosh{\frac{\zeta}{2}} & i\sinh{\frac{\zeta}{2}} & 0 & 0 \\ 
 -i\sinh{\frac{\zeta}{2}} & \cosh{\frac{\zeta}{2}} & 0 & 0 \\ 
 0 & 0 & \cosh{\frac{\zeta}{2}} & -i\sinh{\frac{\zeta}{2}} \\ 
 0 & 0 & i\sinh{\frac{\zeta}{2}} & \cosh{\frac{\zeta}{2}}
 \end{pmatrix}
 \\
 \boldsymbol{\Lambda}_{B3}= \begin{pmatrix}
 e^{-\frac{\zeta}{2}} & 0 & 0 & 0 \\ 
 0 & e^{\frac{\zeta}{2}} & 0 & 0 \\ 
 0 & 0 & e^{\frac{\zeta}{2}} & 0 \\ 
 0 & 0 & 0 & e^{-\frac{\zeta}{2}}
 \end{pmatrix}
 &
 \boldsymbol{\Lambda}_{B4}= \begin{pmatrix}
 \cosh{\frac{\zeta}{2}} & 0 & i\sinh{\frac{\zeta}{2}} & 0 \\ 
 0 & \cosh{\frac{\zeta}{2}} & 0 & i\sinh{\frac{\zeta}{2}} \\ 
 -i\sinh{\frac{\zeta}{2}} & 0 & \cosh{\frac{\zeta}{2}} & 0 \\ 
 0 & -i\sinh{\frac{\zeta}{2}} & 0 & \cosh{\frac{\zeta}{2}} 
 \end{pmatrix}
 \end{aligned}
\end{equation}
as the boosts along all axes with $\zeta$ rapidities, and where from the identity \eqref{comm} in the form $\boldsymbol{\Lambda}\gamma^{\nu}\boldsymbol{\Lambda}^{-1}\Lambda^{\mu}_{\phantom{\mu}\nu}=\gamma^{\mu}$ it is possible to check the form of the matrices $\Lambda^{\mu}_{\phantom{\mu}\nu}$ yielding the expressions for the real representation of the generic transformation of the Lorentz group.

Dirac spinor fields in five dimensions are defined exactly like in the four-dimensional counterpart, and so in terms of a column of four complex scalar fields indicated with $\psi$ in general, although now they will be characterized by different transformation properties, as
it is clear from the fact that, despite all matrices on the left are the standard four-dimensional ones with block-diagonal form, all matrices written on the right are those involving the fifth dimension without block-diagonal form even in chiral representation: as a consequence of this non-reducible structure there will always be some mixing between the left-handed and the right-handed parts, which cannot therefore be defined as separate projections.

Because it is still $\gamma_{0}\gamma_{\mathcal{A}}^{\dagger}\gamma_{0}\!=\!\gamma_{\mathcal{A}}$ for $\mathcal{A}=1,2,3$ but now it is also $\gamma_{0}\gamma_{4}^{\dagger}\gamma_{0}\!=\!\gamma_{4}$ then we have $\gamma_{0}\gamma_{\mu}^{\dagger}\gamma_{0}\!=\!\gamma_{\mu}$ in general: as a consequence $\gamma_{0}S_{\mu\nu}^{\dagger}\gamma_{0}\!=\!-S_{\mu\nu}$ and thus $\gamma_{0} \boldsymbol{\Lambda}^{\dagger}\gamma_{0}\!=\!\boldsymbol{\Lambda}^{-1}$ ensuring that the conjugation is $\bar{\psi} = \psi^{\dagger}\gamma_{0}$ also in $5$ dimensions.

The Dirac spinorial bilinears are defined as
\begin{subequations}
\begin{eqnarray}
&4i\bar{\psi}S^{\mu\nu}\psi= M^{\mu\nu}\\
&\bar{\psi}\gamma^{\mu}\psi= U^{\mu}\\
&\bar{\psi}\psi=\Phi
\end{eqnarray}
\end{subequations}
so to turn out all real and transforming with $\Lambda^{\mu}_{\phantom{\mu}\nu}$ as the real representation of the Lorentz group. They verify
\begin{subequations}
\begin{eqnarray}
&M_{\mu\nu}U^{\mu}=0\label{F1}\\
&\frac{1}{4}M_{\mu\nu}M^{\mu\nu}=U_{\mu}U^{\mu}=\Phi^{2}\label{F2}
\end{eqnarray}
\end{subequations}
as a direct substitution would straightforwardly show and which are called Fierz identities.

The spinorial covariant derivative is defined as in the usual
four-dimensional framework. In particular, equation \eqref{comm} gives
the possibility to see what is the form of the Fock-Ivanenko
coefficients in the spinorial covariant derivative
\begin{equation}
 D_{i}\psi = \partial_{i}\psi + \omega_{i}^{\phantom{i}\mu\nu}S_{\mu\nu}\psi
\end{equation}
defined in this way so that once applied to the gamma matrices it
vanishes identically.
\subsubsection{Polar decomposition}

We now proceed to discuss a way to categorize spinors according to the
idea of the Lounesto classification \cite{L} (a complementary but
similar classification is done by Cavalcanti in reference
\cite{Cavalcanti:2014wia}). In the standard case of four dimensions, we
have two main classes according to whether the scalar and
pseudo-scalar are both zero or not, but in the five-dimensional case
the pseudo-scalar has become the fifth component of the five-vector
$U^{\mu}$ and because there exist transformations mixing it with the
other components then requiring its vanishing does not make sense
covariantly; however, it still makes sense to require the vanishing of
the scalar, which is unchanged: thus we have the two cases, given by
either $\Phi = 0$ or not.

As in references \cite{L,Cavalcanti:2014wia}, we shall call these cases
\emph{singular} when $\Phi = 0$, and \emph{regular} in the most general
circumstance where this constraint does not hold. In the
four-dimensional situation, it has been shown in reference \cite{Fabbri:2016msm}
that it is always possible to find Lorentz transformations bringing
the spinor in the most general case in the following form
\begin{equation}
 \psi = \phi
 \begin{pmatrix}
 e^{i\frac{\beta}{2}}\\
 0\\
 e^{-i\frac{\beta}{2}}\\
 0
 \end{pmatrix}
\end{equation}
where $\phi$ and $\beta$ are real scalars or, in the special case in which
both scalar and pseudo-scalar vanish, in the form
\begin{equation}
 \psi = e^{i\xi} \begin{pmatrix}
 \cos{\frac{\theta}{2}}\\
 0\\
 0\\
 \sin{\frac{\theta}{2}}
 \end{pmatrix}
\end{equation}
where $\xi$ and $\theta$ are real scalars, and should be set to zero, $\xi =
0$, if the spinor is also charged. These results are general, obtained
only through use of the specific form of the Lorentz transformations.

In the five-dimensional situation, Lorentz transformations are changed
and therefore we should expect that even the very same analysis would
furnish somewhat different results on the spinorial structure: we will
now see how this polar decomposition would change.

As above, we split the regular and singular situations in what follows.

In the case of \emph{regular} spinor we have that no constraint is
given: the Fierz orthogonal identity, equation \eqref{F2}, tells us that
$M_{\mu\nu}M^{\mu\nu} = U_{\mu}U^{\mu} > 0$ and thus, in particular, that $U^{\mu}$ is
time-like with the consequence that it is always possible to employ
four boosts to remove one by one all its spatial components; then,
employing rotations we can remove components of the $M^{\mu\nu}$
tensor. Because the fifth component of the five-vector $U^{\mu}$ is what
in the four-dimensional case would be the pseudo-scalar, the analysis
essentially reduces to that of the four-dimensional case plus the
additional restriction of vanishing of the pseudo-scalar. This means
that the most general spinor can always be Lorentz-transformed into
the form
\begin{equation}
 \psi = \phi
 \begin{pmatrix}
 1\\
 0\\
 1\\
 0
 \end{pmatrix}
\end{equation}
in terms of a single scalar function. Thus regular spinors in five
dimensions are simpler than in the four-dimensional case. Not so, and
indeed the opposite, for singular spinors.

For \emph{singular} spinors we have the constraint $\Phi = 0$ identically:
the Fierz identity \eqref{F2} tells that $M_{\mu\nu}M^{\mu\nu} = U_{\mu}U^{\mu} =
0$ and in particular $U^{0}U^{0} - \sum_{\mathcal{A}=1}^3{ U^{\mathcal{A}}U^{\mathcal{A}}} = U^{4}U^{4}
> 0$ so that we can employ three boosts to set $U^{\mathcal{A}}=0$ ($\mathcal{A}=1,2,3$)
identically; we still have all rotations at our disposal to align
$M^{0\mathcal{A}}$ and $M^{\mathcal{A}4}$ along the third axis, so that we get
\begin{equation}
 \psi = \phi
 \begin{pmatrix}
 e^{i\frac{\pi}{4}}\\
 0\\
 e^{-i\frac{\pi}{4}}\\
 0
 \end{pmatrix}
\end{equation}
in terms of a single scalar function. Thus singular spinors in five
dimensions are more complex than in the four-dimensional case, and indeed
they are not singular at all.

So, whereas in four dimensions the real scalar degrees of freedom of
regular spinors are \emph{two} while for singular spinors are
\emph{zero}, in five dimensions the real scalar degrees of freedom for
both regular and singular spinors are just one. As a matter of fact,
it is even possible to write both in the same manner as
\begin{equation}
 \psi = \phi
 \begin{pmatrix}
 e^{i p \frac{\pi}{4}}\\
 0\\
 e^{- i p \frac{\pi}{4}}\\
 0
 \end{pmatrix}
 \label{spinor}
\end{equation}
with $p = 0$ for regular and $p = 1$ for singular spinors.

With the spinor in equation \eqref{spinor}, we get that the spinorial
covariant derivative is
\begin{equation}
 D_{i}\psi = \left( \partial_{i}\ln{\phi}\mathbb{I} + \omega_{i}^{\phantom{i}\mu\nu}S_{\mu\nu}
 \right) \psi
 \label{decspinder}
\end{equation}
so that the spinorial covariant derivative actually acts as a local
matrix operator.
\subsection{Dynamical equations}

We consider Einstein-Cartan gravity coupled to a Dirac field in five
dimensions. The Lagrangian function of the theory is then given by
\begin{equation}
 \mathscr{L} = R + \frac{i}{2}\left(\bar{\psi}\gamma^iD_{i}\psi
 - D_{i}\bar{\psi}\gamma^{i}\psi\right) - m\bar{\psi}\psi
\end{equation}
where $\gamma^i = \gamma^\mu e_\mu^i$ and $m$ is the mass of the spinor field. Upon
variations, the field equations are the Einstein gravitational field
equations
\begin{equation}
 \label{einsteinequations}
 R_{ij} - \frac{1}{2}Rg_{ij} = \Sigma_{ij}
\end{equation}
the Sciama-Kibble torsional field equations
\begin{equation}
 \label{torsionequations}
 T_{tsi} = S_{tsi}
\end{equation}
and the Dirac spinor field equations
\begin{equation}
 \label{diracequations}
 i\gamma^hD_h\psi-m\psi=0
\end{equation}
where $\Sigma_{ij}$ and $S_{ijh}$ are the energy and the spin density
tensors respectively expressed as
\begin{equation}
 \label{4.2a}
 \Sigma_{ij}=\frac{i}{4}\left(\bar{\psi}\gamma_{i}D_{j}\psi-D_{j}\bar{\psi}\gamma_{i}\psi\right)
\end{equation}
and 
\begin{equation}
 \label{4.2b}
 S_{ijh}=\frac{i}{2}\bar{\psi}\left\{\gamma_{h},S_{ij}\right\}\psi
 =-\frac{1}{8}\epsilon_{ijhab}M^{ab}
\end{equation}
in which the complete antisymmetry of the Dirac spin density tensor
clarifies why we considered only a totally antisymmetric torsion
without loss of generality. Moreover, we recall that the Dirac spinor
field equations imply the conservation laws
\begin{eqnarray}
 \nabla_i\Sigma^{ij}= T^{jik}\Sigma_{ik}+\frac{1}{2}S_{pqi}R^{pqij}
 \\
 \nabla_h S^{ijh}= \Sigma^{ji}-\Sigma^{ij}
 \label{antisymm}
\end{eqnarray}
where the antisymmetry of spin and torsion has been
systematically used \cite{FV1}.

Inserting equations \eqref{torsionequations} into the decomposition
\eqref{tensordecomposition}, we can rewrite
equation \eqref{einsteinequations} in the form
\begin{equation}
 \label{4.7}
 \tilde{R}_{ij}-\frac{1}{2}\tilde{R} g_{ij}+\frac{1}{2} \tilde{\nabla}_p S_{ji}^{\;\;\;p}-\frac{1}{4}S_{pi}^{\;\;\;q} S_{jq}^{\;\;\;p}+\frac{1}{8} S_{qpr} S^{qpr}g_{ij} = \Sigma_{ij}
\end{equation}
and it is an easy matter to verify that the antisymmetric part of
equation \eqref{4.7} amounts to the conservation laws \eqref{antisymm}.
Therefore the significant part of equation \eqref{4.7} reduces to the
symmetric one and reads as
\begin{equation}
 \label{4.12}
 \tilde{R}_{ij}-\frac{1}{2}\tilde{R} g_{ij}
 =\frac{i}{4}\left[\bar{\psi}\gamma_{(i}\tilde{D}_{j)}\psi
 -\tilde{D}_{(j}\bar{\psi}\gamma_{i)}\psi\right]+\frac{3}{32}\Phi^{2}g_{ij}
\end{equation}
where $\tilde{D}_i$ denotes the spinorial covariant derivative induced
by the Levi-Civita connection.

In detail, equations \eqref{4.12} are deduced by making use of the identities
\begin{subequations}
 \label{4.9.0}
 \begin{gather}
 \label{4.9}
 \Sigma_{ij} = \frac{i}{4}\left[\bar\psi\gamma_i\tilde{D}_j\psi-(\tilde{D}_j\bar\psi)\gamma_i\psi\right] -\frac{1}{4}S_{pi}^{\;\;\;q} S_{jq}^{\;\;\;p}
 \\
 \label{4.10}
 S_{hi}^{\;\;\;p}S_{jp}^{\;\;\;h} =
 \frac{1}{64}(\bar\psi[\gamma^\alpha,\gamma^\beta]\psi)(\bar\psi[\gamma_\alpha,\gamma_\beta]\psi)g_{ij}-\frac{1}{32}(\bar\psi[\gamma_i,\gamma_p]\psi)(\bar\psi[\gamma_j,\gamma^p]\psi)
 \\
 \label{4.11}
 S_{hqp}S^{hqp} = \frac{3}{32}(\bar\psi[\gamma^\alpha,\gamma^\beta]\psi)(\bar\psi\gamma_\alpha\gamma_\beta\psi)
 \\
 \label{4.11bis}
 \left(\bar\psi\left[\gamma_\alpha,\gamma_\beta\right]\psi\right)\left[\gamma^\alpha,\gamma^\beta\right]\psi
 =-16\left(\bar\psi\psi\right)\psi
 \\
 \label{4.11tris}
 \left(\bar\psi\left[\gamma_\alpha,\gamma_\beta\right]\psi\right)\left(\bar\psi\left[\gamma^\alpha,\gamma^\beta\right]\psi\right) =-16\left(\bar\psi\psi\right)^2
 \end{gather}
\end{subequations}
Analogously, Dirac equation, i.e. equation \eqref{diracequations}, can be
worked out by using the decomposition
\begin{equation}
 \label{4.8trisa}
 D_i \psi = \tilde{D}_i \psi+\frac{1}{8}T_{ijh}\gamma^h\gamma^j\psi
\end{equation}
and then expressed in the final form
\begin{equation}
 i \gamma^h \tilde{D}_h \psi - \left( \frac{3}{8}\Phi + m \right) \psi=0
\end{equation}
where the non-linearity has been translated into a simple correction
to the mass term.
\subsubsection{Polar equations}
By employing the polar form of the spinor field, it is possible to provide a corresponding polar form of the spinor field equations: this has been done in \cite{h1}, where all Gordon decompositions of the polar form of Dirac field equations were found, and in \cite{Fabbri:2016laz}, where we isolated the sub-set of Gordon decompositions that imply the polar form of Dirac field equations. Therefore, the combined results of \cite{h1,Fabbri:2016laz} can be used to see that the Dirac field equations in polar form can equivalently be written in terms of two real vector field equations.

When the same procedure is done in the $5$ dimensional case however, we obtain a single real vector equation
\begin{equation}
\partial_{s}\Phi+\tilde{\omega}^{h}_{\phantom{h}hs}\Phi
+\frac{1}{2}\epsilon_{hijps}\tilde{\omega}^{hij}U^{p}=0
\end{equation}
giving first-order derivatives of the only degree of freedom in terms of the spin connection.

Therefore, in the four-dimensional case, the Dirac spinor equations, accounting for $8$ real equations, can be decomposed in two real vector equations, accounting for a corresponding number of $8$ real equations, while in the five-dimensional case, the Dirac spinor equations, accounting for $8$ real equations, can be decomposed in one real vector equation, accounting for $5$ real equations, with $3$ Dirac equations converting into constraints.
\section{Geometrical reduction to the four-dimensional spacetime}
\subsection{Break-down of the penta-dimensional structure into the
 tetra-dimensional spacetime plus a trivial dimension}

We refer the manifold $\pi : Q\to M$ to local fiber coordinates $x^{A},x^4$ (with capital Latin letters running from zero to three), where $x^{A}$ are coordinates on $M$. Local fiber coordinate transformations are of the form
\begin{equation}
\label{coordinatetransformations}
\left\{
\begin{aligned}
&\bar{x}^{A}= \bar{x}^{A}(x^{B})\\
&\bar{x}^4= \bar{x}^4 (x^{B},x^4)
\end{aligned}
\right.
\end{equation}
and in particular, whenever the subset of $1$-forms $e^\Psi=e^\Psi_{A}(x^{B})\,dx^{A}$ (with capital Greek letters running from zero to three) results to be the pull-back of a tetrads field on $M$, then the quantity
\begin{equation}
\label{1.000}
\pi^*(\bar g) := g+e^4\otimes e^4
\end{equation} 
defines a metric tensor $\bar g$ on $M$. Such a construction is invariant under the action of the sub-group of $SO(1,4)$ consisting of the matrices of the form
\begin{equation}
 \label{sottogruppoLorentz}
 \Lambda^\Psi_{\;\;\Phi} (\pi(x))=
 \begin{pmatrix}
 \Lambda^{\Psi}_{\;\;{\Phi}} (\pi(x)) & 0 
 \\
 0 & 1
 \end{pmatrix}
\end{equation}
with $\Lambda^{\Psi}_{\;\;{\Phi}} (\pi(x)) \in SO (1,3)$, $\forall x\in Q$. 

Under simultaneous coordinate and Lorentz transformations,
equations \eqref{coordinatetransformations} and \eqref{sottogruppoLorentz},
the spin-connection coefficients undergo the transformation laws
\begin{subequations}
 \label{spinconnectiontrasformazione}
 \begin{gather}
 \bar{\omega}_A^{\;\;\Psi\Phi} = \Lambda^\Psi_{\;\;\Sigma}\Lambda^\Phi_{\;\;\Omega}\de{x^B}/de{\bar{x}^A}\omega_B^{\;\;\Sigma\Omega}-\Lambda_\Sigma^{\;\;\Omega}\de{\Lambda^\Psi_{\;\;\Omega}}/de{x^B}\de{x^B}/de{\bar{x}^A}\eta^{\Sigma\Phi}+\Lambda^\Psi_{\;\;\Sigma}\Lambda^\Phi_{\;\;\Omega}\de x^4/de{\bar{x}^A}\omega_4^{\;\;\Sigma\Omega}\\
 \bar{\omega}_A^{\;\;\Psi 4} = \Lambda^\Psi_{\;\;\Phi}\de{x^B}/de{\bar{x}^A}\omega_B^{\;\;\Phi 4}+\Lambda^\Psi_{\;\;\Phi}\de x^4/de{\bar{x}^A}\omega_4^{\;\;\Phi 4}\\
 \bar{\omega}_4^{\;\;\Psi 4} = \Lambda^\Psi_{\;\;\Phi}\de x^4/de{\bar{x}^4}\omega_4^{\;\;\Phi 4}\\
 \bar{\omega}_4^{\;\;\Psi\Phi} = \Lambda^\Psi_{\;\;\Sigma}\Lambda^\Phi_{\;\;\Omega}\de x^4/de{\bar{x}^4}\omega_4^{\;\;\Sigma\Omega}
 \end{gather} 
\end{subequations}
showing that, limited to the transformation subgroups
\eqref{coordinatetransformations} and \eqref{sottogruppoLorentz}, a
spin-connection $\omega_A^{\;\;\Psi\Phi}$ on $M$ can be always lifted to a
corresponding spin-connection on $Q$ by setting $\omega_A^{\;\;\Psi 4} =
\omega_4^{\;\;\Psi 4} = \omega_4^{\;\;\Psi\Phi} =0$. Conversely, given a spin-connection
$\omega_i^{\;\;\mu\nu}$ on $Q$ (independent of the $x^4$ coordinate), the
subset of coefficients $\omega_A^{\;\;\Psi\Phi}$ defines a spin-connection on $M$
provided that the group of coordinate transformations is reduced to
\begin{equation}
 \label{reducedcoordinatetransformations}
 \left\{
 \begin{aligned}
 \bar{x}^{A}= \bar{x}^{A}(x^{B})
 \\
 \bar{x}^4= \bar{x}^4 (x^4)
 \end{aligned}
 \right.
\end{equation}
consistent with the trivial fibration $Q=M\times U$ being $U$ a generic
one-dimensional manifold.
\subsection{The prototypical case of the Bianchi-I models}
Let us consider a Bianchi type I metric of the form
\begin{equation}
 \label{3.1}
 ds^2=dt^2-a^2(t)\,dx^2-b^2(t)\,dy^2-c^2(t)\,dz^2-d^2(t)\,du^2
\end{equation}
identifying $x^0 =t$, $x^1 =x$, $x^2 =y$, $x^3 =z$ and $x^4= u$ for
simplicity; the components of the f\"{u}nfbein associated with the
metric \eqref{3.1} are expressed as
\begin{equation}
 e^\mu_0=\delta^\mu_0, \quad e^\mu_1=a(t) \delta^\mu_1, \quad e^\mu_2=b(t) \delta^\mu_2, \quad e^\mu_3=c(t) \delta^\mu_3, \quad e^\mu_4=d(t) \delta^\mu_4, 
\end{equation}
with
\begin{equation}
 e^0_\mu=\delta^0_\mu, \quad e^1_\mu=\frac{1}{a(t)}\delta^1_\mu, \quad e^2_\mu=\frac{1}{b(t)}\delta^2_\mu, \quad e^3_\mu=\frac{1}{c(t)}\delta^3_\mu \quad e^4_\mu=\frac{1}{d(t)}\delta^4_\mu,
\end{equation}
where $\mu =0,1,2,3,4$ as in general. The non-null components of the Levi-Civita connection associated with the metric \eqref{3.1} are
\begin{equation}
 \begin{aligned}
 \tilde{\Gamma}_{10}^{\;\;\;1}& = \frac{\dot a}{a}, &
 \tilde{\Gamma}_{20}^{\;\;\;2}& = \frac{\dot b}{b}, &
 \tilde{\Gamma}_{30}^{\;\;\;3}& = \frac{\dot c}{c}, &
 \tilde{\Gamma}_{40}^{\;\;\;4}& = \frac{\dot d}{d}
 \\
 \tilde{\Gamma}_{11}^{\;\;\;0}& = a{\dot a}, &
 \tilde{\Gamma}_{22}^{\;\;\;0}& = b{\dot b}, &
 \tilde{\Gamma}_{33}^{\;\;\;0}& = c{\dot c}, &
 \tilde{\Gamma}_{44}^{\;\;\;0}& = d{\dot d}
 \end{aligned}
\end{equation}
and consequently we also have that
\begin{equation}
 \tilde{\Omega}_1=\frac{1}{2}{\dot a}\gamma^1\gamma^0, \quad \tilde{\Omega}_2=\frac{1}{2}{\dot b}\gamma^2\gamma^0, \quad \tilde{\Omega}_3=\frac{1}{2}{\dot c}\gamma^3\gamma^0, \quad \tilde{\Omega}_4=\frac{1}{2}{\dot d}\gamma^4\gamma^0
\end{equation} 
are the non-zero coefficients of the spinorial connection needed to construct the spinorial covariant derivative in the form $\tilde{D}_i\psi=\partial_i\psi-\tilde{\Omega}_i\psi$.

With the spinorial covariant derivative and \eqref{F2} the Dirac spinor field equations are
\begin{subequations}\label{3.16}
 \begin{equation}\label{3.16a}
 \dot\psi+\frac{\dot\tau}{2\tau}\psi+im\gamma^0\psi+\frac{3i}{8} (\bar{\psi}\psi)\gamma^0\psi	=0
 \end{equation}
 \begin{equation}\label{3.16b}
 \dot{\bar{\psi}}+\frac{\dot\tau}{2\tau}\bar{\psi}-im\bar{\psi}\gamma^0-\frac{3i}{8} (\bar{\psi}\psi)\bar{\psi}\gamma^0=0
 \end{equation}
\end{subequations}
where we have defined $\tau = abcd$ as the volume element. Multiplying equation \eqref{3.16a} by $\bar{\psi}$ and equation \eqref{3.16b} by $\psi$ and summing the results, we obtain 
\begin{equation}\label{3.17}
 \frac{d}{dt} (\tau\bar{\psi}\psi)=0
\end{equation}
yielding
\begin{equation}\label{3.20}
 \bar{\psi}\psi=\frac{C}{\tau}
\end{equation}
where $C$ is a suitable integration constant.

Analogously, evaluating the Einstein equations \eqref{4.12} for the metric \eqref{3.1} and using again equation \eqref{F2}, we get 
\begin{subequations}
 \label{3.10}
 \begin{equation}
 \label{3.10a}
 \frac{\dot a}{a}\frac{\dot b}{b}+\frac{\dot b}{b}\frac{\dot c}{c}+\frac{\dot a}{a}\frac{\dot c}{c}+\frac{\dot a}{a}\frac{\dot d}{d}+\frac{\dot b}{b}\frac{\dot d}{d}+\frac{\dot c}{c}\frac{\dot d}{d}=+\frac{m}{2}\bar{\psi}\psi + \frac{9}{32}\left(\bar{\psi}\psi\right)^2
 \end{equation}
 \begin{equation}
 \label{3.10b}
 \frac{\ddot b}{b}+\frac{\ddot c}{c}+\frac{\ddot d}{d}+\frac{\dot b}{b}\frac{\dot c}{c}+\frac{\dot b}{b}\frac{\dot d}{d}+\frac{\dot c}{c}\frac{\dot d}{d}= \frac{3}{32}\left(\bar{\psi}\psi\right)^2
 \end{equation}
 \begin{equation}
 \label{3.10c}
 \frac{\ddot a}{a}+\frac{\ddot c}{c}+\frac{\ddot d}{d}+\frac{\dot a}{a}\frac{\dot c}{c}+\frac{\dot a}{a}\frac{\dot d}{d}+\frac{\dot c}{c}\frac{\dot d}{d}= \frac{3}{32}\left(\bar{\psi}\psi\right)^2
 \end{equation}
 \begin{equation}
 \label{3.10d}
 \frac{\ddot a}{a}+\frac{\ddot b}{b}+\frac{\ddot d}{d}+\frac{\dot a}{a}\frac{\dot b}{b}+\frac{\dot a}{a}\frac{\dot d}{d}+\frac{\dot b}{b}\frac{\dot d}{d}= \frac{3}{32}\left(\bar{\psi}\psi\right)^2
 \end{equation}
 \begin{equation}
 \label{3.10e}
 \frac{\ddot a}{a}+\frac{\ddot b}{b}+\frac{\ddot c}{c}+\frac{\dot a}{a}\frac{\dot b}{b}+\frac{\dot b}{b}\frac{\dot c}{c}+\frac{\dot a}{a}\frac{\dot c}{c}= \frac{3}{32}\left(\bar{\psi}\psi\right)^2
 \end{equation}
\end{subequations}
together with the conditions
\begin{subequations}
 \label{3.11}
 \begin{equation}
 \label{3.11a}
 \Sigma_{(12)}=0\quad \Rightarrow \quad a \dot{b}-b \dot{a}=0 \quad \cup \quad \bar{\psi}\gamma^4\gamma^3\psi =0
 \end{equation}
 \begin{equation}
 \label{3.11b}
 \Sigma_{(23)}=0\quad \Rightarrow \quad c \dot{b}-b \dot{c}=0 \quad \cup \quad \bar{\psi}\gamma^4\gamma^1\psi =0
 \end{equation}
 \begin{equation}
 \label{3.11c}
 \Sigma_{(13)}=0\quad \Rightarrow \quad a \dot{c}-c \dot{a}=0 \quad \cup \quad \bar{\psi}\gamma^4\gamma^2\psi =0
 \end{equation}
 \begin{equation}
 \label{3.11d}
 \Sigma_{(14)}=0\quad \Rightarrow \quad a \dot{d}-d \dot{a} =0 \quad \cup \quad \bar{\psi}\gamma^2\gamma^3\psi =0 
 \end{equation}
 \begin{equation}
 \label{3.11e}
 \Sigma_{(24)}=0\quad \Rightarrow \quad b \dot{d}-d \dot{b} =0 \quad \cup \quad \bar{\psi}\gamma^1\gamma^3\psi =0 
 \end{equation}
 \begin{equation}
 \label{3.11f}
 \Sigma_{(34)}=0\quad \Rightarrow \quad c \dot{d}-d \dot{c} =0 \quad \cup \quad \bar{\psi}\gamma^1\gamma^2\psi =0
 \end{equation}
\end{subequations}
with equations $\Sigma_{0{\cal A}}=0$ (${\cal A}=1,2,3,4$) being automatically satisfied identities.

Subtracting equations \eqref{3.10b} from \eqref{3.10c}, \eqref{3.10c} from \eqref{3.10d} and \eqref{3.10d} from \eqref{3.10e}, we obtain the relations
\begin{subequations}
 \label{3.12}
 \begin{equation}
 \label{3.12a}
 \frac{a}{b}=X_1 e^{Y_1\int{\frac{dt}{\tau}}}
 \end{equation}
 \begin{equation}
 \label{3.12b}
 \frac{b}{c}=X_2 e^{Y_2\int{\frac{dt}{\tau}}}
 \end{equation}
 \begin{equation}
 \label{3.12c}
 \frac{c}{d}=X_3 e^{Y_3\int{\frac{dt}{\tau}}}
 \end{equation}
\end{subequations}
where $X$ and $Y$ are suitable integration constants. A linear
combination of equations \eqref{3.10a}-\eqref{3.10e} gives
\begin{equation}\label{3.21}
3\ddot\tau/\tau=2m\bar{\psi}\psi + \frac{3}{2}\left(\bar\psi\psi\right)^2
\end{equation}
Together with equations \eqref{3.12}, the equation \eqref{3.21} accounts
for the dynamics of the metric \eqref{3.1}, while equation
\eqref{3.10a} plays the role of a constraint on the initial data.
\subsubsection{Compatibility with polar form}
In the previous section we have established the fact that even a most general spinor field can always be reduced, by employing Lorentz transformations down to the polar form \eqref{spinor}. In this section we have studied specific types of universes with metric \eqref{3.1}: despite the polar form can always be achieved, the Dirac spinor is a field, so the needed Lorentz transformation is local, and this in general produces additional contributions in the f\"{u}nfbein, which means that choosing the f\"{u}nfbein \emph{cannot} be done in general.

Since we have done so, it is necessary at this point to spend time to check whether or not our choices are all compatible. And if yes, what other constraints can be implemented.

Our goal is to make sure that the polar form be compatible with restrictions \eqref{3.11}. We consider now the standard representation of the gamma matrices so that, writing the generic spinor according to the expression
\begin{equation}
 \psi=\begin{pmatrix}
 \psi_1\\
 \psi_2\\
 \psi_3\\
 \psi_4
 \end{pmatrix},
\end{equation}
the restrictions become
\begin{subequations}
 \begin{equation}
 \label{3.11abis}
 \bar{\psi}\gamma^4\gamma^3\psi =0 \quad \Longleftrightarrow \quad
 \left(-\psi_1^*\psi_1+\psi^*_2\psi_2-\psi^*_3\psi_3+\psi^*_4\psi_4\right)
 = 0,
 \end{equation}
 \begin{equation}
 \label{3.11bbis}
 \bar{\psi}\gamma^4\gamma^1\psi =0 \quad \Longleftrightarrow \quad
 \left(\psi_1^*\psi_2+\psi^*_2\psi_1+\psi^*_3\psi_4+\psi^*_4\psi_3\right)
 = 0,
 \end{equation}
 \begin{equation}
 \label{3.11cbis}
 \bar{\psi}\gamma^4\gamma^2\psi =0 \quad \Longleftrightarrow \quad
 \left(\psi_1^*\psi_2-\psi^*_2\psi_1+\psi^*_3\psi_4-\psi^*_4\psi_3\right)
 = 0,
 \end{equation}
 \begin{equation}
 \label{3.11dbis}
 \bar{\psi}\gamma^2\gamma^3\psi =0 \quad \Longleftrightarrow \quad
 \left(\psi_1^*\psi_2+\psi^*_2\psi_1-\psi^*_3\psi_4-\psi^*_4\psi_3\right)
 = 0,
 \end{equation}
 \begin{equation}
 \label{3.11ebis}
 \bar{\psi}\gamma^1\gamma^3\psi =0 \quad \Longleftrightarrow \quad
 \left(\psi_1^*\psi_2-\psi^*_2\psi_1-\psi^*_3\psi_4+\psi^*_4\psi_3\right)
 = 0,
 \end{equation}
 \begin{equation}
 \label{3.11fbis}
 \bar{\psi}\gamma^1\gamma^2\psi =0 \quad \Longleftrightarrow \quad
 \left(-\psi_1^*\psi_1+\psi^*_2\psi_2+\psi^*_3\psi_3-\psi^*_4\psi_4\right)
 = 0,
 \end{equation}
\end{subequations}
It seems physically meaningful requiring that the scale factor of the
fourth spatial dimension differs from the other ones. This means,
therefore, that constraints in equations \eqref{3.11dbis}-\eqref{3.11fbis}
have to be always imposed.

General solutions of Dirac equations \eqref{3.16} are given in the form
\begin{equation}
 \label{general_solutions}
 \psi=\frac{1}{\sqrt{2\tau}}
 \begin{pmatrix}
 A_1 e^{-i\left(mt+\frac{3C}{8}\int{\frac{dt}{\tau}}\right)}\\
 A_2 e^{-i\left(mt+\frac{3C}{8}\int{\frac{dt}{\tau}}\right)}\\
 A_3 e^{i\left(mt+\frac{3C}{8}\int{\frac{dt}{\tau}}\right)}\\
 A_4 e^{i\left(mt+\frac{3C}{8}\int{\frac{dt}{\tau}}\right)}
 \end{pmatrix},
\end{equation} 
where the $A_{j}=r_{j}e^{i\theta_{j}}$ with $j=1,2,3,4$ are four
complex integration constants. Equations
\eqref{3.11dbis}-\eqref{3.11fbis} yield
\begin{eqnarray}
 r_1r_2\cos(\theta_2-\theta_1)= r_3r_4\cos(\theta_4-\theta_3),
 \\
 r_1r_2\sin(\theta_2-\theta_1)= r_3r_4\sin(\theta_4-\theta_3),
 \\
 r_2^2+r_3^2= r_1^2+r_4^2,
\end{eqnarray}
eventually giving either
\begin{equation}
 \label{general_solutions_3}
 \begin{pmatrix}
 A_1\\
 A_2\\
 A_3\\
 A_4
 \end{pmatrix}
 =
 \begin{pmatrix}
 re^{i\theta}\\
 0\\
 re^{i\varphi}\\
 0
 \end{pmatrix}
 \quad
 \text{or}
 \quad
 \begin{pmatrix}
 A_1\\
 A_2\\
 A_3\\
 A_4
 \end{pmatrix} =
 \begin{pmatrix}
 0\\
 re^{i\theta}\\
 0\\
 re^{i\varphi}
 \end{pmatrix},
\end{equation}
where $r\geq 0$, $\theta$ and $\varphi$ are arbitrary real numbers,
and also
\begin{equation}
 \label{general_solutions_2}
 \begin{pmatrix}
 A_1\\
 A_2\\
 A_3\\
 A_4
 \end{pmatrix} =
 \begin{pmatrix}
 qe^{i\theta_1}\\
 re^{i\theta_2}\\
 qe^{i\theta_3}\\
 re^{i\theta_4}
 \end{pmatrix},
\end{equation}
where $q\geq 0$ and $r\geq 0$ are arbitrary, while the phases satisfy
$\theta_2-\theta_1=\theta_4-\theta_3+2k\pi$ being $k$ an integer
number. However, in any case, all the admissible solutions of Dirac
equations \eqref{3.16} with constraints
\eqref{3.11dbis}-\eqref{3.11fbis} satisfy $\Phi=0$ necessarily.

We must now check compatibility with the polar form. The condition
$\Phi=0$ implies that we are treating singular spinor fields, and for
these, Fierz identities in equations \eqref{F1} and \eqref{F2} tell
\begin{gather}
 M_{ik}U^{i}=0\\
 M_{ab}M^{ab} = U_{a}U^{a} = 0
\end{gather}
identically; we have already discussed how in this situation it is always possible to boost into the \emph{rest frame}, the one for which $U^{A}=0$ identically, and in this frame Fierz identities
\begin{gather}
 M_{40}=0\\
 M^{0A}U^{0}=M^{4A}U^{4}\\
 2M_{0A}M^{0A}+M_{AB}M^{AB}+2M_{4A}M^{4A} = 0\\
 U^{0}U^{0} = U^{4}U^{4}
\end{gather}
can be further re-arranged into
\begin{gather}
 |M^{0A}|=|M^{4A}|\\
 |U^{0}| = |U^{4}|\\
 M_{40}=0\\
 M^{AB} = 0\label{dipolespatial}
\end{gather}
identically in this frame. Notice that equation \eqref{dipolespatial} is equivalent to requiring the validity of constraints \eqref{3.11dbis}-\eqref{3.11fbis} which we know should always be imposed. 

This establishes the compatibility of the isotropy constraints with the polar form.
\section{Special cases}
\subsection{Isotropic case}
In the maximally isotropic case $a=b=c$ the constraints
\eqref{3.11abis}-\eqref{3.11cbis} are automatically satisfied. The
solution of Dirac spinor field equations satisfying the remaining
constraints \eqref{3.11dbis}-\eqref{3.11fbis} is
given by equation \eqref{general_solutions} together with
\eqref{general_solutions_3} and \eqref{general_solutions_2}, which
imply the further condition $\bar{\psi}\psi =0$ ($C=0$). This simplify the
Einstein-like equations. In particular equation \eqref{3.21} reduces to $\ddot\tau=0$. Therefore, we distinguish $\tau =\beta$ and $\tau=\alpha t+\beta$ as the only two
admissible sub-cases.

If $\tau=\beta$ from equations \eqref{3.12}, we deduce
\begin{equation}
 a(t)=\left(\beta X\right)^{\frac{1}{4}}e^{\frac{Y}{4\beta}t}
\end{equation}
and
\begin{equation}
 d(t)=\beta^{\frac{1}{4}}X^{-\frac{3}{4}}e^{-\frac{3Y}{4\beta}t}
\end{equation}
from which we have
\begin{equation}
 \frac{\dot a}{a}=\frac{Y}{4\beta} \qquad \text{and} \qquad \frac{\dot d}{d}=-\frac{3Y}{4\beta}
\end{equation}
which can be inserted into \eqref{3.10a} giving $Y=0$ necessarily, which amounts to a five-dimensional flat spacetime filled by a constant Dirac field. This is an unphysical solution. 

If $\tau=\alpha t+\beta$, we have
\begin{equation}
 a(t)=X^{\frac{1}{4}}\left(\alpha t+\beta\right)^{\frac{Y+1}{4}}
\end{equation}
and
\begin{equation}
 d(t)=X^{-\frac{3}{4}}\left(\alpha t+\beta\right)^{\frac{-3Y+1}{4}}
\end{equation}
implying
\begin{equation}
 \frac{\dot a}{a}=\frac{\alpha\left(Y+1\right)}{4\left(\alpha t+\beta\right)} \qquad {\rm and} \qquad \frac{\dot d}{d} =\frac{\alpha\left(-3Y+1\right)}{4\left(\alpha t+\beta\right)}
\end{equation}
and it is a straightforward matter to see that $Y=1$ and $Y=-1$ are the only initial data consistent with \eqref{3.10a}: for $Y=-1$ the scale factor $a$ is constant, while the scale factor $d$ expands, so the resulting cosmological scenario is not acceptable. For $Y=1$, $d$ contracts and $a$ undergoes a decelerated expansion, the associated four-dimensional spacetime could describe a Friedmann era of our Universe, but it can represent neither a Universe undergoing inflation nor a Universe characterized by a dark energy era.
\subsection{Partially isotropic case}
In the partially isotropic case as for instance if $a=b\neq c$, we
have that also constraints \eqref{3.11bbis} and \eqref{3.11cbis} have
to be imposed. It is easy to see that the admissible solutions for the
Dirac field are now given by equations \eqref{general_solutions} and
\eqref{general_solutions_3}.

If $\tau =\beta$, from equations \eqref{3.12} and renaming some
integration constants, we get the identities
\begin{eqnarray}
a(t)=b(t)=X_1\left(\frac{\beta}{X_1^2X_2}\right)^{\frac{1}{4}}e^{\left(\frac{1}{2}Y_1-\frac{1}{4}Y_2\right)t} \\
c(t)= X_2\left(\frac{\beta}{X_1^2X_2}\right)^{\frac{1}{4}}e^{\left(-\frac{1}{2}Y_1+\frac{3}{4}Y_2\right)t}\\
d(t)= \left(\frac{\beta}{X_1^2X_2}\right)^{\frac{1}{4}}e^{\left(-\frac{1}{2}Y_1-\frac{1}{4}Y_2\right)t}
\end{eqnarray}
which imply the relations
\begin{equation}
\frac{\dot a}{a}=\frac{\dot b}{b}=\frac{1}{2}Y_1-\frac{1}{4}Y_2, \quad \frac{\dot c}{c}=-\frac{1}{2}Y_1+\frac{3}{4}Y_2 \quad{\rm and} \quad \frac{\dot d}{d}=-\frac{1}{2}Y_1-\frac{1}{4}Y_2.
\end{equation}
Inserting these into \eqref{3.10a} it is easy to see that the only
admissible solution corresponds to $Y_1 =Y_2 =0$ representing a
five-dimensional flat spacetime.

If $\tau=\alpha t+\beta$ the factor scales of the metric \eqref{3.1}
are expressed as
\begin{eqnarray}
a(t)=b(t)= X_1\left(\frac{1}{X_1^2X_2}\right)^{\frac{1}{4}}\left(\alpha t+\beta\right)^{\frac{1}{4}\left(1+2Y_1-Y_2\right)}\\
c(t)= X_2\left(\frac{1}{X_1^2X_2}\right)^{\frac{1}{4}}\left(\alpha t+\beta\right)^{\frac{1}{4}\left(1-2Y_1+3Y_2\right)}\\
d(t)= \left(\frac{1}{X_1^2X_2}\right)^{\frac{1}{4}}\left(\alpha t+\beta\right)^{\frac{1}{4}\left(1-2Y_1-Y_2\right)}
\end{eqnarray}
and therefore we have the relations
\begin{gather}
 \frac{\dot a}{a}=\frac{\dot b}{b}=\frac{\alpha\left(1+2Y_1-Y_2\right)}{4\left(\alpha t+\beta\right)}
 \\
 \frac{\dot c}{c}=\frac{\alpha\left(1-2Y_1+3Y_2\right)}{4\left(\alpha t+\beta\right)}
 \\
 \frac{\dot d}{d}=\frac{\alpha\left(1-2Y_1-Y_2\right)}{4\left(\alpha t+\beta\right)},
\end{gather}
which can be inserted into equation \eqref{3.10a} yielding a constraint for
the initial data
\begin{equation}
 \label{constr}
 4Y_1^2-4Y_1Y_2-3-4Y_2+3Y_2^2 = 0.
\end{equation}
In order to obtain accelerated expansion for the scale factors $a(t)$
and $c(t)$ as well as contraction for $d(t)$ we should have solutions
of equation \eqref{constr} satisfying the conditions
\begin{equation}
 \label{cond}
 1+2Y_1-Y_2>4, \quad 1-2Y_1+3Y_2>4, \quad {\rm and} \quad 1-2Y_1-Y_2<0,
\end{equation}
but once again it is straightforward to verify that
equation \eqref{constr}, subjected to the constraints \eqref{cond}, has no
solutions (indeed, the system of equations \eqref{cond} admits solutions
only for $Y_2>3$ while equation \eqref{constr} possesses real solutions
only for
$\frac{2-\sqrt{10}}{2}\leq Y_2 \leq \frac{2+\sqrt{10}}{2} <3$).

Instead, if we require that the scale factors $a(t)$ and $c(t)$ expand
but not necessarily accelerating, namely if we only impose
\begin{equation}
 \label{4.2.11}
 1+2Y_1-Y_2>0, \quad 1-2Y_1+3Y_2>0, \quad {\rm and} \quad 1-2Y_1-Y_2<0,
\end{equation}
it is easily seen that solutions do exist. 

For example, the pair $Y_2=1$ and $Y_1=\frac{1+\sqrt{5}}{2}$ solves
simultaneously equations \eqref{constr} and \eqref{4.2.11}. 
\subsection{Totally anisotropic case}
In the totally anisotropic situation we have $a\neq b$, $a\neq c$, $b\neq c$ and therefore also the constraint \eqref{3.11abis} have to be imposed. However, solutions \eqref{general_solutions} and 
\eqref{general_solutions_3} are clearly not compatible with \eqref{3.11abis}. Therefore, the totally anisotropic case is not viable.
\section{Conclusion}
In this paper, we have considered a Dirac spinor field in a five-dimensional background, and proceeded to reduce such set-up to a four-dimensional spacetime. We have seen that it is always possible to choose a frame in which the spinor field can be written-without loss of generality-in the form shown in equation \eqref{spinor}, so that we also have $\Phi = \bar{\psi}\psi = 2\phi^{2}(1 - p)$ with $p=0$ and $p=1$ designing regular and singular spinors respectively. The corresponding decomposition of the Dirac spinor field equation is
\begin{equation}
\partial_{s}\ln{\phi} + \frac{1}{2}\tilde{\omega}^{h}_{\phantom{h}hs}
+\frac{1}{4}\epsilon_{hijps}\tilde{\omega}^{hij}U^{p}/\Phi=0,
\end{equation}
giving first-order derivatives of $\phi$ in terms of the spin connection, and reducing to
\begin{equation}
\epsilon_{hijps}\tilde{\omega}^{hij}U^{p}=0
\end{equation}
as a constraint over the spin connection in the case of singular spinor fields. Therefore, a four-dimensional spinor field obtained as a reduction of a five-dimensional spinor field differs from a genuine four-dimensional spinor field for the fact that the former has only one degree of freedom, the module, while the latter has in general two degrees of freedom, the module and the so-called Yvon-Takabayashi angle. Additionally, in the former case the number of spinor field equations (that is $8$) does not match the number of field equations after the polar decomposition (which are only $5$). As a consequence of this circumstance, we conclude that an initially five-dimensional
spinor field later reduced to a four-dimensional spinor field is always more constrained than a genuine four-dimensional spinor field.

Eventually, we have studied the case of five-dimensional geometry of Bianchi-I anisotropic Universes, investigating whether a contraction of the fourth spatial scale factor can give rise to a possibly accelerated expansion of the reduced four-dimensional Universe. Due to the further constraints that the spinor field have to satisfy in five-dimensions, we found that in the maximally and partially isotropic cases, the reduced four-dimensional Universe can-at the most-experience a Friedmann expansion, but accelerated phases are not allowed. In the totally anisotropic case, solutions do not exist. Once more we want to highlight that, the behaviour of
anisotropic universes defined in five dimensions and later reduced to four dimensions, is different from what is obtained for anisotropic universes in $4$ dimensions \cite{Saha1,Saha3,Saha2}.

The underlying theory, as well as the example of the Bianchi-I Universes, indicates that an initially five-dimensional Universe with spinors later reduced to a four-dimensional Universe with spinors is more constrained than a genuine four-dimensional Universe with spinors.

\end{document}